\documentclass[EPiCempty]{easychair}
\usepackage[utf8]{inputenc}
\usepackage{doc}
\usepackage{todonotes}
\usepackage{wrapfig}
\usepackage{url}
\usepackage{tabularx}
\newcolumntype{Y}{>{\centering\arraybackslash}X}
\usepackage[binary-units,per-mode=symbol]{siunitx}
\usepackage[absolute,showboxes]{textpos}
\usepackage{pgfplots}
\pgfplotsset{compat=newest,
  /pgfplots/ybar legend/.style={
    /pgfplots/legend image code/.code={%
      \draw[##1,/tikz/.cd,yshift=-0.25em](0cm,0cm) rectangle (5pt,0.8em);
    },
  },
}
\usepgfplotslibrary{units}
\makeatletter
\pgfplotsset{
  unit code/.code 2 args=
    \begingroup
    \protected@edef\x{\endgroup\si{#2}}\x
}
\makeatother
\usetikzlibrary{pgfplots.groupplots}
\usetikzlibrary{patterns}
\definecolor{corered}{RGB}{200,0,0}
\definecolor{coredarkgray}{RGB}{51,51,51}
\definecolor{coreblue}{RGB}{0, 46, 125}

\newcommand{\mytitle}{DoS Protection through Credit Based Metering - Simulation-Based Evaluation for Time-Sensitive Networking in Cars}

\title{\mytitle}

\author{
Philipp Meyer
\and
Timo Häckel
\and
Franz Korf
\and
Thomas C. Schmidt
}

\institute{
  Department of Computer Science \\
  Hamburg University of Applied Sciences, Germany
  \\
  \email{\{philipp.meyer, timo.haeckel, franz.Korf, t.schmidt\}@haw-hamburg.de}
 }

\begin{document}

\authorrunning{Meyer, Häckel, Korf and Schmidt}


\titlerunning{DoS Protection through Credit Based Metering}

\maketitle

\setlength{\TPHorizModule}{\paperwidth}
\setlength{\TPVertModule}{\paperheight}
\TPMargin{5pt}
\begin{textblock}{0.8}(0.1,0.02)
    \noindent
    \footnotesize
    If you cite this paper, please use the original reference:
    P. Meyer, T. H{\"a}ckel, F. Korf, and T.~C. Schmidt. DoS Protection through Credit Based Metering - Simulation-Based Evaluation for Time-Sensitive Networking in Cars. In: \emph{Proceedings of the 6th International OMNeT++ Community Summit}. September, 2019, Easychair.
\end{textblock}

\begin{abstract}
Ethernet is the most promising solution to reduce complexity and enhance the bandwidth in the next generation in-car networks.
Dedicated Ethernet protocols enable the real-time aspects in such networks. One promising candidate is the IEEE 802.1Q Time-Sensitive Networking protocol suite.
Common Ethernet technologies, however, increases the vulnerability of the car infrastructure as they widen the attack surface for many components. In this paper proposes an IEEE 802.1Qci based algorithm that on the one hand, protects against DoS attacks by metering incoming Ethernet frames. On the other hand, it adapts to the behavior of the Credit Based Shaping algorithm, which was standardized for Audio/Video Bridging, the predecessor of Time-Sensitive Networking. A simulation of this proposed Credit Based Metering algorithm evaluates the concept.
\end{abstract}

\vspace{-3mm}
\section{Introduction}
In today's vehicles, a multitude of sensors, actors, and electronic control units (ECUs) are used to enable enhanced performance, comfort, and safety through advanced driver assistance systems.
Even autonomous driving will be realized in future generations.
These additions result in complex communication over different proprietary bus technologies in multiple domains.

Ethernet technologies are used to set up efficient and straightforward communication in future generations.
Real-time Ethernet protocols enable the compliance of communication requirements and enhance the reliability of Standard Ethernet.
Promising candidates are the Time-Sensitive Networking (TSN) protocols by the IEEE (\url{https://1.ieee802.org/tsn/}).

The main focus of those protocols is Quality of Service guarantees.
The integration of future cars in the IoT context opens its systems to global communication.
These online capabilities and the domain interconnection increases the attack surface of safety-critical functions like brakes and the motor control units.
Attacks could manipulate driving characteristics and could provide fatal consequences for vehicle and passengers.

Therefore, security has to be an essential goal for the development of the next generations of on-board communications technologies.
The TSN standard IEEE 802.1Qci addresses some security concerns by filtering ingress traffic on network node ports.

This work provides a simulation-based evaluation of IEEE 802.1Qci with a Credit Based Meter (CBM) algorithm concept. This concept enforces the reserved bandwidth of a stream and is one solution to protect network nodes from Denial of Service (DoS) attacks.
IEEE 802.1Qci and the CBM concept are implemented in the OMNeT++ environment to enable evaluations.

This paper is organized as follows: Section \ref{sec:background} presents previous and related work.
In Section \ref{sec:cbm}, the developed simulation environment, and the credit based metering are shown.
Section \ref{sec:case_study} presents a case study followed by an evaluation of the implemented simulation models and concepts.
The paper closes with a conclusion and outlook in section \ref{sec:conclusion}.

\vspace{-3mm}
\section{Background \& Related Work}
\label{sec:background}

\begin{wrapfigure}{R}{0.55\textwidth}
\centering
\vspace{-6mm}
\includegraphics[width=0.55\textwidth]{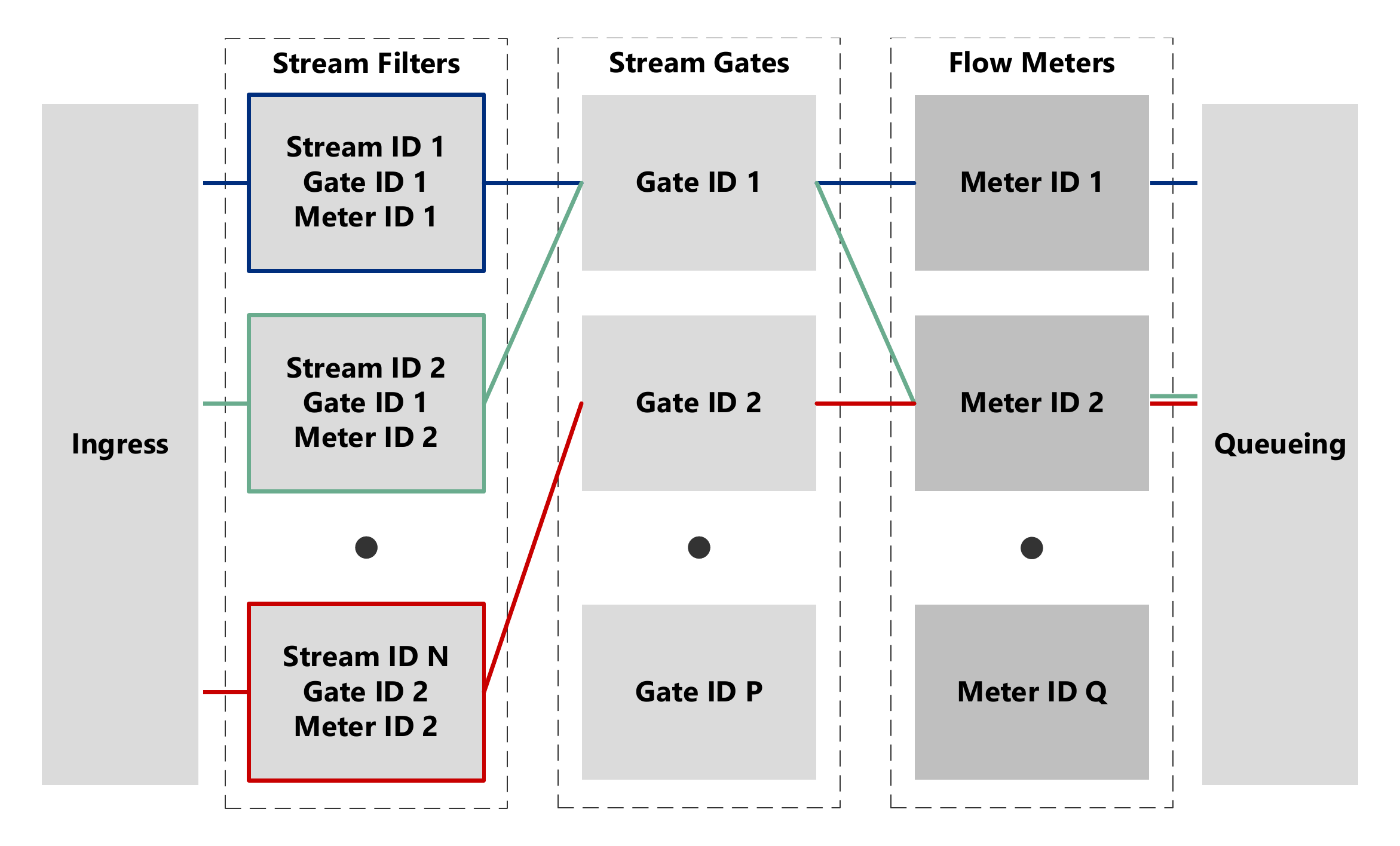}
\vspace{-6mm}
\caption{IEEE 802.1Qci per-stream filtering and policing}
\label{fig:TSN_Ingress}
\end{wrapfigure}

The on-board network of a vehicle is a highly distributed system defined by its electronic control units (ECUs).
At present, proprietary bus technologies (CAN, MOST, FlexRay) enable the communication of control units.

Future development will lead to a stepwise transition towards flat, switched Ethernet networks \cite{mk-ae-15}.
Such networks have to support simultaneous transmission of messages with different priorities to maintain safe time-critical communication.
Real-time Ethernet protocols are used to guarantee the different quality of service classes.

The Time-Sensitive Networking (TSN) \cite{ieee8021q-18} real-time Ethernet standard is a collection of protocols which adapts to network requirements of cyber-physical systems.
Domains for this protocol suite are, for example, industrial control facilities or in-car networks.

The focus in this work is the sub-standard IEEE 802.1Qci \cite{ieee8021qci-17}.
It describes an ingress control through per-stream filtering and policing.
Figure \ref{fig:TSN_Ingress} shows the structure of the filtering and policing specification.
The shown mechanism has an instance after each port ingress of a TSN networking device. The result is that all ingress traffic is filtered.
There are three stages an incoming frame has to pass through before it is queued.
The first stage consists of a set of stream filters.
They configure which gates and meters are responsible for handling frames of a specific stream id.
Secondly, there are stream gates.
Those have one of the two states "OPEN" or "CLOSED".
This state can change based on a static defined schedule based on a system-wide clock.
If the gate is "CLOSED", the frame will be dropped.
If the responsible gate is in the state "OPEN", the frame will be handled by the responsible flow meter.
The flow meters stage contain unique algorithms to assert if a message is allowed.
After the meter allows a frame, it gets queued in the network node for subsequent forwarding or processing.
This work presents a meter concept called Credit Based Meter (CBM) (See section \ref{sec:cbm}).

The importance of security measures for in-car networks is shown in various related work \cite{iii-sauJR-18}, \cite{rwpvm-panJR-17}.
A fundamental work is from Checkoway et al. \cite{cmkas-ceaas-11}.
They examine interfaces that are part of the attack surface in a car.
These interfaces are classified in three categories: Physical access (ODB-II, CD, USB), short distance wireless access (Bluetooth, WiFi, Remote-Keyless-Entry) and long distance wireless access (GPS, digital radio, mobile services).
The authors gained access to the on-board network in each category using reverse engineering and debugging.

Miller and Valasek \cite{mv-reupv-15}  described in detail how to obtain control over an unaltered passenger vehicle.
They gained control over safety-critical elements like engines and brakes over a remote cellular connection of the infotainment system.
The infotainment system is part of the internal CAN-based communication infrastructure, and CAN buses in this infrastructure contain virtually no security measures.
At this point, the authors used reverse engineering to get information over the communication between in-car control units.
In the next step CAN messages are forwarded over the infotainment system into the internal communication which is received and processed by the control units.

In consequence, security measures must be included in future automotive communication systems. Simulations of in-car networks \cite{sdkks-eifre-11} are an essential method to study the behavior of such systems in detail.
This work presents one solution for protecting the in-car communication from DoS attacks and analyses it in the simulator.
In those scenarios, a compromised stream is used to burst frames into networking devices to overload there capabilities. The result could be lost or delayed frames of not compromised streams.

\vspace{-3mm}
\section{Credit Based Metering}
\label{sec:cbm}

\begin{wrapfigure}{R}{0.55\textwidth}
\centering
\vspace{-8mm}
\includegraphics[width=0.55\textwidth]{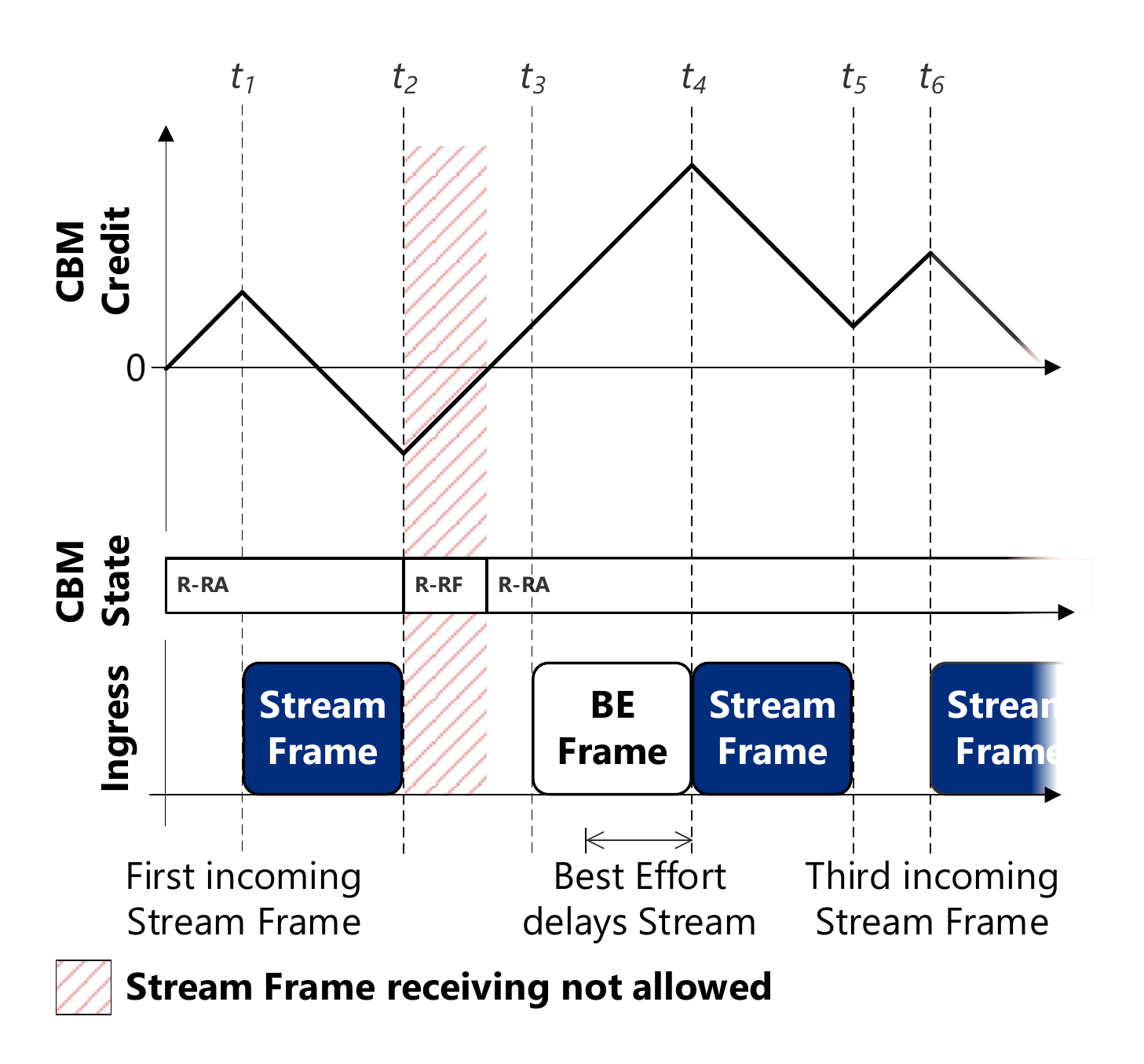}
\vspace{-8mm}
\caption{Credit Based Metering state machine}
\label{fig:CBM_Credit}
\end{wrapfigure}


The Credit Based Meter (CBM) is an ingress counterpart for the Credit Based Shaper (CBS) egress behavior defined in IEEE 802.1Qbv \cite{ieee8021qbv-16}.
In general, a CBM is based on a credit value manipulated by two different gradients called "$idleslope$" and "$sendslope$".
Frame reception is allowed when the credit is greater or equal to zero.
When the credit is lower than zero, an incoming frame will be discarded.

\vspace{-5mm}
\begin{align}
  \label{equ:IDLESLOPE}
  idleslope &= RB\\
  \label{equ:SENDSLOPE}
  sendslope &= RB - B
\end{align}


These gradients are composed of an accumulated reserved bandwidth ($RB$) of the streams passing the meter and a total bandwidth ($B$) of the port (See equation \ref{equ:IDLESLOPE} and \ref{equ:SENDSLOPE}).

Additionally, the CBM contains a maximum burst size parameter ($Burst_{max}$) configuring the maximum count of frames that are allowed in an incoming stream burst.
This is used in combination with the sending duration ($T_{duration}$) of one frame. $T_{duration}$ is composed of the frame size ($FS_{stream}$), the port bandwidth ($B$) and the Ethernet inter frame gap ($T_{ifg}$) to calculate the maximum credit value ($Credit_{max}$) of the CBM shown in equations \ref{equ:DURATION} and \ref{equ:MAXCREDIT}.

\begin{equation}
  \label{equ:DURATION}
  T_{duration} = \frac{FS_{stream}}{B} + T_{ifg}
\end{equation}

\begin{equation}
  \label{equ:MAXCREDIT}
  Credit_{max} = |sendslope| * T_{duration} * (Burst_{max} - 1)
\end{equation}

Because a burst of one frame is allowed when the credit is 0 $Burst_{max}$ has to be substracted by one. So the definition of $Burst_{max} = 1$ results in $Credit_{max} = 0$.

The CBM has two states. They are "RUNNING RECEIVING ALLOWED" (R-RA) and "RUNNING RECEIVING FORBIDDEN" (R-RF).
When the CBM starts the state is R-RA and the credit is set to zero.
The credit starts to increases according to $idleslope$ till the first frame is incoming or the credit reaches the maximum ($Credit_{max}$).

In the R-RA state, the credit is decreased by "$sendslope$" for the receiving duration of a frame.
When the frame is queued the credit increases again with "$idleslope$". If the credit is greater or equal to zero, a new frame reception is allowed, and the credit decreases again by "$sendslope$".
When the credit reaches the maximum, it stays on this value until a frame is incoming.

If the credit is lower than 0, the state will be switched to R-RF.
In R-RF each incoming frame will be deleted. Simultaneously, the credit increases with "$idleslope$".
The state is changed back to R-RA when the credit reaches 0.

In figure \ref{fig:CBM_Credit}, an example of the CBM algorithm behavior is shown.
Firstly, the state is R-RA and the credit is 0 and increases according to "$idleslope$" until the first frame arrives (see $t_1$ in figure \ref{fig:CBM_Credit}).
The credit decreases by "$sendslope$" for the duration of the frame ($t_2$ in figure \ref{fig:CBM_Credit}).
Now the state is changed to R-RF and the credit increases by "$idleslope$" till it reaches 0.
The state changes to R-RA, and the credit increases further until the next frame arrives.
This is delayed by an incoming best effort (BE) frame ($t_3$ in figure \ref{fig:CBM_Credit}).
The next frame arrives and the credit is decreased again ($t_4$ to $t_5$ in figure \ref{fig:CBM_Credit}).
The credit increases till the third frame receiving starts ($t_6$ in figure \ref{fig:CBM_Credit}). So again the credit decreases by "$sendslope$" until the end of the transmission duration.

The performance of the CBM is dependent on $Burst_{max}$. A target configuration of this parameter is as low as possible and still supports a valid worst-case scenario.
On the one side, this is because of the counterpart CBS. The valid maximum frame burst of a stream that is produced by a CBS algorithm egress is dependent on its specific worst-case scenario.
On the other side, an attack creating a maximum frame burst could not harm the network because it is designed to support the worst-case traffic workload.

There are different ways to determine a minimal $Burst_{max}$ value.
One example is analyzing the worst-case burst behavior for each streams output port ($Burst_{out}$).
$Burst_{max}$ has to be calculated, as shown in equation \ref{equ:MAXBURST} to allow one closeup frame following the burst.

\begin{equation}
  \label{equ:MAXBURST}
  Burst_{max} = Burst_{out} + 1
\end{equation}

Another example of determining a $Burst_{max}$ value is by simulating different configurations to find one that fits the requirements.

\vspace{-3mm}
\section{Case Study}
\label{sec:case_study}
This section evaluates the integration of the Credit Based Meter algorithm inside IEEE 802.1Qci.
This is done by using the OMNeT++ (\url{https://omnetpp.org/}) simulation environment with INET (\url{https://inet.omnetpp.org/}) and  our CoRE4INET framework.
CoRE4INET enables in-car network simulations \cite{mkss-smcin-19} and the simulations of TSN features \cite{msks-eatts-13}.
For this work, the CoRE4INET is extended with IEEE 802.1Qci and CBM implementations.

\begin{figure}[h]
\centering
\includegraphics[width=1\columnwidth]{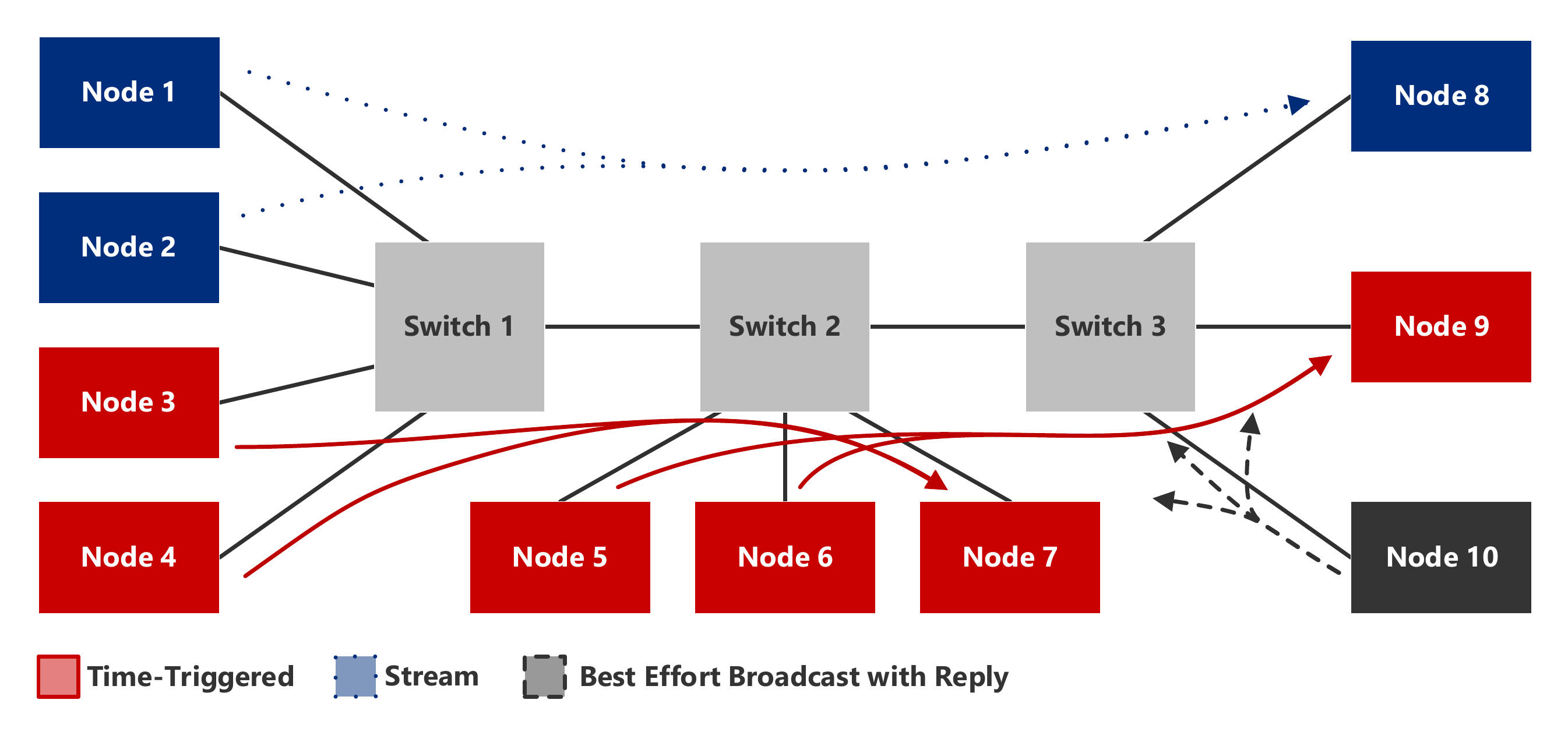}
\caption{Simulation Topology}
\label{fig:SIM_Topology}
\end{figure}

The chosen topology is known from previous work \cite{msks-eatts-13} and is designed to create critical links with multiple concurrent traffic.
Figure \ref{fig:SIM_Topology} shows this topology.
In this topology, time-triggered traffic is based on TDMA with the highest priority.
Two configurations are simulated.
The first is a configuration with active CBM filtering.
The second one emplaces a compromised "Node 1" into the simulation, which is spamming a DoS attack into the network.

For both simulations the base configuration is as follows:
\begin{itemize}
  \item All links are configured with a bandwidth of \SI{100}{\mega\bit\per\second}.
  \item "Node 1" and "Node 2" are the sources of "Stream 1" and "Stream 2" with "Node 8" as its destination. Both streams have a reserved route with an individual bandwidth of \SI{25}{\mega\bit\per\second}. Each stream is passing a CBM on all devices and gates are "OPEN".
  \item Full-size time-triggered frames are generated by "Node 3", "Node 4", "Node 5" and "Node 6". The first two are received by "Node 7" and the latter by "Node 9". In all switches, a gap of \SI{123}{\micro\second} is configured to allow intermediate frame bursts.
  \item For extra background traffic "Node 10" is broadcasting full-size best-effort Ethernet frames. All nodes are replying by sending a full-size best-effort frame back to "Node 10".
\end{itemize}

The worst-case output stream burst sizes ($Burst_{out}$) are known in this base configuration.
They are 2 for "Node 1" and "Node 2" and, because of the concurrent TDMA traffic 4 for "Switch 1" and "Switch 2".
Therefore the $Burst_{max}$ value for CBM filtering in "Switch 1" is 3 for both input ports and 5 for the input metering in "Switch 2" and "Switch 3" (See equation \ref{equ:MAXBURST} in section \ref{sec:cbm}).


\begin{figure}[t]
\centering
\begin{minipage}[b]{.45\textwidth}
\centering
\begin{tikzpicture}
\begin{axis}[width=1\textwidth,
  change x base,
  x SI prefix=micro, x unit=\micro\second,
  xlabel=End-to-end latency,
  ylabel=No. of frames,
  legend pos=north east,
  ybar,
  xmin=0.00035,
  xmax=0.00095,
  ymin=0,
  ymax=20000,
  ]
\addplot[fill,
  coreblue,
  draw opacity=0,
  fill opacity=0.6,
  ] table [x, y, col sep=comma] {data/filtering/hist_end_to_end_latency_str1.csv};
\addplot[fill,
  coredarkgray,
  postaction={pattern color=white, pattern=north west lines},
  draw opacity=0,
  fill opacity=0.6,
  ] table [x, y, col sep=comma] {data/filtering/hist_end_to_end_latency_str2.csv};
\legend{Stream 1, Stream 2}
\end{axis}
\end{tikzpicture}
\vspace{-3mm}
\caption{End-to-end latency of frames per stream without attack}
\label{fig:hist_filtering}
\end{minipage}
\hspace{2pt}
\begin{minipage}[b]{.45\textwidth}
\centering
\begin{tikzpicture}
\begin{axis}[width=1\textwidth,
  change x base,
  x SI prefix=micro, x unit=\micro\second,
  xlabel=End-to-end latency,
  ylabel=No. of frames,
  yticklabel pos=right,
  legend pos=north east,
  ybar,
  xmin=0.00035,
  xmax=0.00095,
  ymin=0,
  ymax=20000,
  ]
\addplot[fill,
  coreblue,
  draw opacity=0,
  fill opacity=0.6,
  ] table [x, y, col sep=comma] {data/attack/hist_end_to_end_latency_str1.csv};
\addplot[fill,
  coredarkgray,
  postaction={pattern color=white, pattern=north west lines},
  draw opacity=0,
  fill opacity=0.6,
  ] table [x, y, col sep=comma] {data/attack/hist_end_to_end_latency_str2.csv};
\legend{Stream 1, Stream 2}
\end{axis}
\end{tikzpicture}
\vspace{-3mm}
\caption{End-to-end latency of frames per stream during an attack}
\label{fig:hist_attack}
\vspace{-0.5mm}
\end{minipage}
\end{figure}

\begin{wrapfigure}{R}{0.6\textwidth}
\vspace{-3mm}
\begin{tikzpicture}
\begin{axis}[width=0.55\textwidth,
  axis y line*=left,
  change x base,
  x SI prefix=mega, x unit=\mega\bit\per\second,
  change y base,
  y SI prefix=mega, y unit=\mega\bit\per\second,
  xlabel=Input bandwidth,
  ylabel=Output bandwidth,
  ymax=60000000,
  ]
\addplot[mark=o,
  coreblue,
  thick,
  ] table [x, y , col sep=comma] {data/attack/spam_interval_study_bandwidth_s1_p1_str1.csv};
\label{plot_1}
\end{axis}
\begin{axis}[width=0.55\textwidth,
  axis x line=none,
  axis y line*=right,
  ylabel=No. of frames dropped per second,
  legend pos=north west,
  legend cell align={left},
  ]
\addlegendimage{/pgfplots/refstyle=plot_1}\addlegendentry{Bandwidth relationship}
\addplot[mark=x,
  coredarkgray,
  thick,
  ] table [x, y , col sep=comma] {data/attack/spam_interval_study_dropped_frames_s1_p1_str1.csv};
\addlegendentry{Frames dropped}
\end{axis}
\end{tikzpicture}
\vspace{-6mm}
\caption{Impact of CBM on Stream 1 in Switch 1}
\label{fig:impact_casestudy_attack}
\end{wrapfigure}
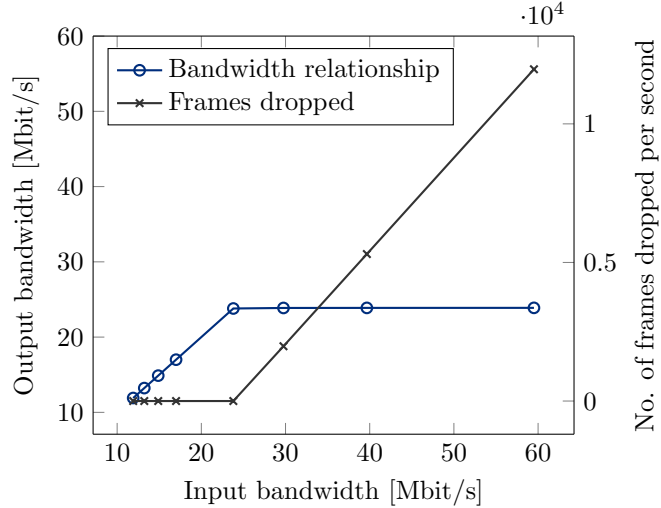

The results shown in this section are a selection of results generated by the simulations.
All shown simulation results are based on 10 seconds duration runs.

The first result set (Figure \ref{fig:hist_filtering} and \ref{fig:hist_attack}) presents and compares the end-to-end latency of the streams in both configurations.
Due to the assumption that valid packets are not influenced by the CBM, these latencies are expected to be nearly the same.

The end-to-end latency of both streams in the configuration with CBM filtering is shown in figure \ref{fig:hist_filtering}.
In this configuration, each node and switch is using a CBM ingress control on each port and for each stream.
The two histograms show the number of frames that arrived at the target with a specific consolidated end-to-end latency.
Blue shows these results for "Stream 1" and grey for "Stream 2".

Figure \ref{fig:hist_attack} shows the end-to-end latency of the streams in a configuration where "Node 1" is corrupted.
All nodes and switches are using CBM ingress control again.
The difference is that "Node 1" is generating the "Stream 1" packets in an invalid pattern.
This is done by spamming subsequent frames.

The comparison shows no significant differences between the configurations.
This demonstrates that CBM is successfully enforcing the correct behavior.
This is done by removing all "Stream 1" frames of the corrupted source that would excel the reserved bandwidth.
Therefore "Stream 2" is not affected by "Node 1" spamming.

\begin{wrapfigure}{R}{0.6\textwidth}
\vspace{-8mm}
\begin{tikzpicture}
\begin{groupplot}[
  group style={
        group name=my plots,
        group size=1 by 3,
        xlabels at=edge bottom,
        xticklabels at=edge bottom,
        vertical sep=2pt
    },
    change x base,
    x SI prefix=milli, x unit=\milli\second,
    xticklabel style={/pgf/number format/fixed, /pgf/number format/precision=3},
    xtick distance=0.000125,
    xlabel=Simulation time,
    xmin=0.1465, xmax=0.147,
    ylabel absolute,
    xmajorgrids,
    ]
  \nextgroupplot[
    width=0.55\textwidth,
    ylabel=Credit value,
    ytick distance=1000,
    ]
  \addplot[corered,sharp plot,update limits=false] coordinates {(0.1465,4707) (0.147,4707)};
           \node[corered] at (axis cs:0.146625,4400) {$Credit_{max}$};
  \addplot[mark=no,
    coreblue,
    thick,
    ] table [x, y, col sep=comma] {data/filtering/vec_credit_s1_str1.csv};
  \draw[ultra thin] (axis cs:\pgfkeysvalueof{/pgfplots/xmin},0) -- (axis cs:\pgfkeysvalueof{/pgfplots/xmax},0);
  \nextgroupplot[
    width=0.55\textwidth,
    height=3cm,
    ylabel style={align=center},
    ylabel={Frame\\on\\ingress},
    ytick distance=1,
    yticklabels={,,},
    ]
  \addplot[mark=no,
    coreblue,
    thick,
    ] table [x, y, col sep=comma] {data/filtering/vec_frames_s1_str1.csv};
  \nextgroupplot[
    width=0.55\textwidth,
    height=5cm,
    y unit=\mega\bit\per\second,
    ylabel=Bandwidth,
    ytick distance=10,
    ]
  \addplot[mark=no,
    coreblue,
    thick,
    ] table [x, y, col sep=comma] {data/filtering/vec_bandwidth_125us_s1_str1.csv};
\end{groupplot}
\end{tikzpicture}
\vspace{-6mm}
\caption{Section of CBM credit, frame, and bandwidth}
\label{fig:section_credit_filtering}
\end{wrapfigure}
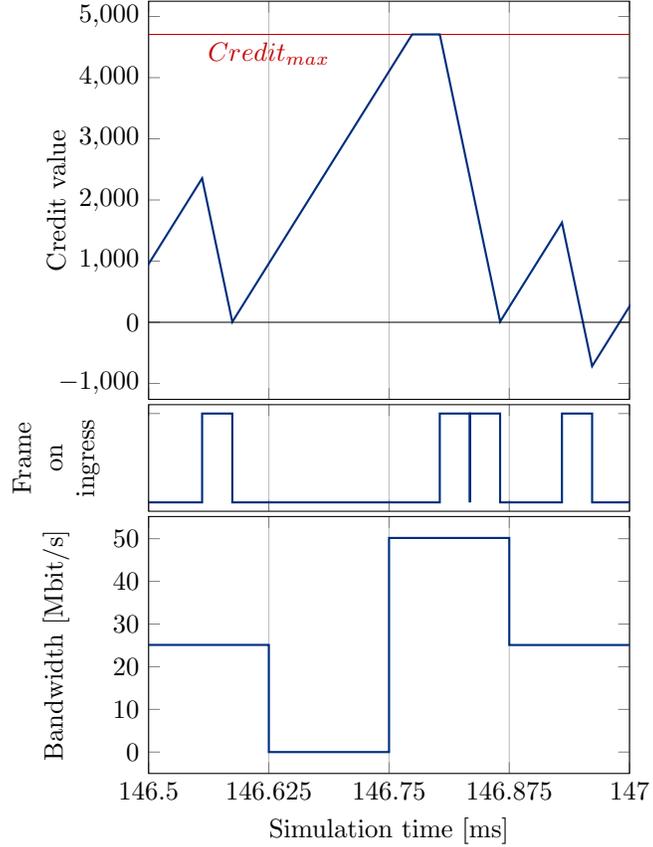

Figure \ref{fig:impact_casestudy_attack} presents output bandwidth size and number of frames dropped in the CBM in "Switch 1" for "Stream 1" produced by eight simulation runs.
For each run, the input bandwidth produced by "Node 1" is incremented.
The reserved bandwidth of \SI{25}{\mega\bit\per\second} is fixed.
It is expected that the output would not cross this reserved bandwidth value.

The result reflects the wanted CBM behavior.
No frame is dropped, and the output bandwidth is the same as the input bandwidth until the input size overshoots the reserved bandwidth of \SI{25}{\mega\bit\per\second}.
At this point, the number of frames dropped increases as a function of the input bandwidth. Because each frame, which would exceed the reserved bandwidth, will be dropped by the CBM.

A selected section of this CBM algorithm is shown in Figure \ref{fig:section_credit_filtering}. It presents the credit value, frame ingress, and output bandwidth for a specific timeslot of the simulation.
"Node 1" produces a valid "Stream 1" packet flow of \SI{25}{\mega\bit\per\second} in this scenario.
Although the CBM output bandwidth never exceeds the reserved bandwidth over time, it allows short crossings like its counterpart CBS.

Because no frame is received between \SI{146.625}{\milli\second} and \SI{146.75}{\milli\second} the credit increases as a function of this duration.
This continues until it reaches its maximum, which is dependent on $Burst_{max}$.

In this case, the $Burst_{max}$ value is 3.
This results in a $Credit_{max}$ value of ca. 4650. The corresponding equation \ref{equ:MAXCREDIT_SIM} shows the calculation of this $Credit_{max}$ value.

Next, a continuous burst of 3 frames would be allowed.
In this case, just two subsequent packages are incoming.
This results in a zoomed in bandwidth of \SI{50}{\mega\bit\per\second} between \SI{146.75}{\milli\second} and \SI{146.875}{\milli\second}.

This shows that the reserved bandwidth could be overshoot massively for shorter periods.
This is dependent on the configuration of $Burst_{max}$.
From this also follows that $Burst_{max}$ value has no influence on the over-time bandwidth restriction.
Buffer sizes have to support the $Burst_{max}$ values to guarantee that they did not overflow.

The CBM enforces this upper barrier.
For a configured network, the maximum latencies could be calculated and are valid even if a malfunction or attack results in an invalid behavior of individual network participants.
This protects the integrity and availability of the in-car communication system.

\begin{align}
  \label{equ:MAXCREDIT_SIM}
  \begin{split}
    Credit_{max} &= |sendslope| * T_{duration} * (Burst_{max} - 1)\\
                 &\approx \SI{75}{\mega\bit\per\second} * \SI{31}{\micro\second} * (3 - 1) = 4650
  \end{split}
\end{align}

\vspace{-3mm}
\section{Conclusion \& Outlook}
\label{sec:conclusion}
The demand for interconnecting an increasing multitude of sensors, actors, and ECUs in today's vehicles guides in-car networks to adapt real-time Ethernet technologies.
Flattening the network in this way creates new vulnerabilities within the in-car network.
The CBM is a solution for a TSN meter algorithm to protect the system against DoS attacks.
It protects the integrity and availability of an in-car communication system by individually controlling the stream input on each ingress port of the network.
The CBM allows all valid traffic patterns of a CBS algorithm.
An attacker could use the burst behavior to shortly overcome the reserved bandwidth restrictions, but the credit boundary limits the bandwidth over an extended period.
This limit is the same as the reserved bandwidth.
The maximum burst parameter has to be as low as possible to gain the best performance.
However, it still must allow the valid worst-case scenario of a specific input port.
This trade-off between performance and worst-case estimation has to be considered.

In future work, the compatibility with other TSN traffic shaper concepts will be evaluated. Furthermore combined operation of different ingress control mechanisms will be simulated.
In addition, the benefits of the ingress control metrics for anomaly detection will be analyzed.

Our simulation models and the extensions in this work are published at \url{sim.core-rg.de}.

\vspace{-3mm}
\section*{Acknowledgments}
\label{sec:acknowledgements}
This work is funded by the German Federal Ministry of Education and Research (BMBF) within the SecVI project.
\vspace{-3mm}
\bibliographystyle{plain}
\bibliography{special,HTML-export/all_generated}

\end{document}